\documentclass[aps,prl,twocolumn,groupedaddress,showpacs]{revtex4}

\usepackage{graphicx}
\usepackage{amsmath}
\usepackage{amssymb}
\usepackage{dcolumn}
\usepackage{epsf} 

\begin{document}


\title{Tunneling Density of States of the Interacting 
Two--Dimensional Electron Gas}

\author{J\"org Rollb\"uhler}
\author{Hermann Grabert}
\affiliation{Physikalisches Institut,
             Albert-Ludwigs-Universit\"at,
             Hermann-Herder-Stra{\ss}e 3,
             D--79104 Freiburg, Germany}

\date{April 17, 2003}

\begin{abstract}
We investigate the influence of electron--electron interactions
on the density of states of a ballistic two--dimensional 
electron gas. The density of states is determined 
nonperturbatively by means of path integral techniques allowing 
for reliable results near the Fermi surface, 
where perturbation theory breaks down. 
We find that the density of states is suppressed at the Fermi 
level to a finite value. This suppression factor grows with 
decreasing electron density and is weakened by the presence of gates.
\end{abstract}


\maketitle

The suppression of electron tunneling into a conductor at 
low bias voltages, a phenomenon known as zero-bias anomaly (ZBA), has 
been under consideration theoretically and 
experimentally for more than 20 years 
[1--8].
This effect can be related to a reduction of the 
(tunneling) density of states (DOS) at the Fermi level, which 
is a clear signature of interaction effects that are otherwise 
often disguised in good conductors.

In the past the focus has been mainly on disordered systems with
slow diffusive electron motion enhancing Coulomb effects
so that strong ZBAs arise \cite{AltshulerAronov85}.
More recently, attention has shifted to one-dimensional ballistic 
wires \cite{Bockrath99etc}, 
where the Tomonaga-Luttinger model leads to a power 
law suppression of the DOS \cite{Fisher97}. Similar to 
disorder, the reduction of dimension hinders fast spreading and 
forces the particles to interact more strongly.
Two-dimensional ballistic systems are a 
borderline case between strongly interacting one-dimensional fermions
and weakly interacting three-dimensional systems. The DOS is also of 
special interest, since two-dimensional electron gases (2DEGs) 
arise in a variety of semiconductor microstructures, such as
GaAs heterostructures and Si inversion layers 
that have attracted a lot of attention lately. 
Further, 2DEGs provide
electrodes in experimental tunneling setups measuring properties
of 1D quantum wires. Since such experiments do not test the 
DOS of the wire alone, it is important to know the form of the
DOS of the electrode.

In earlier work \cite{Khveshchenko98,MA01}, 
based on perturbation theory in the Coulomb interaction, the 
change of the DOS of 2DEGs in the absence of a screening gate
was found to show a cusp 
$\delta \nu(\epsilon)/\nu_{0} = |\epsilon|/4 \epsilon_{\rm F}$
at the Fermi edge.
This prediction is rather irritating since it is independent
of the strength of the Coulomb interaction.
Here we reexamine the problem employing nonperturbative path 
integral techniques and show that the suppression of the DOS at 
the Fermi level sensitively depends on the electron density 
and screening by gates. 
In contrast to the 1D case, in the absence of disorder, the DOS at 
the Fermi level remains finite also at zero temperature yet reduced 
from the bare density 
that is approached at larger energies.
The suppression increases with the effective interaction strength,
i.e., with decreasing electron density.

The interacting electron gas can be described by the 
action for a fermion field $\psi$ coupled to an electric 
potential $\phi$ 
\begin{eqnarray}
 S[\psi^{*}, \psi, \phi] &=& \int_{\cal C} dt \biggl\{
  - \frac{e^{2}}{2} \int \frac{d^{2} q}{(2 \pi)^{2}} \; 
  \phi(-{\bf q}) V_{0}^{-1}({\bf q}) \phi({\bf q}) 
  \nonumber \\
 &+& e \int \frac{d^{2} p}{(2 \pi)^{2}} \frac{d^{2} p'}{(2 \pi)^{2}} \; 
  \bar{\psi}({\bf p}) \phi({\bf p}-{\bf p}') \psi({\bf p}') \biggr\} 
  \nonumber \\ 
 &+& \int \frac{d^{2} p}{(2 \pi)^{2}} \; \bar{\psi}({\bf p})
   G_{0}^{-1}({\bf p}) \psi({\bf p}) \; , \nonumber
\end{eqnarray}
where $V_{0}$ is the Coulomb interaction potential and the operator 
$G_{0}^{-1}({\bf p},t) = i \partial_{t} - \epsilon({\bf p})$ 
(we set $\hbar=1$) contains
the electronic dispersion relation $\epsilon({\bf p})=p^{2}/2m^{*}$ 
with the effective mass $m^{*}$. 
The time integration path is along the Keldysh contour. 
For simplicity we restrict ourselves to spinless Fermions
but account for spin degeneracy by appropriate factors of 2.
Further, we add a source term $\bar{J} \psi + \bar{\psi} J$
to the action allowing us to calculate the Green function later.

The Fermion fields can be integrated out in the standard way \cite{Kopietz97} 
yielding an effective electromagnetic action
\begin{eqnarray}\label{eq:eff_action}
 S_{\rm eff}[\phi] &=& \int_{\cal C} dt \biggl\{
  - \frac{e^{2}}{2} \int \frac{d^{2} q}{(2 \pi)^{2}} \; \phi(-{\bf q}) 
  V_{0}^{-1}({\bf q}) \phi({\bf q}) 
\nonumber \\
 && \hspace{-0.0cm} 
  + {\rm Tr} \ln 
   \left[G_{0}^{-1}({\bf p}) \delta({\bf p}-{\bf p}')  
   + e \phi({\bf p}-{\bf p}') \right] 
\nonumber \\ && \hspace{-0.0cm} 
 + \int \frac{d^{2} p}{(2 \pi)^{2}} \;  
  \bar{J}({\bf p}) G[\phi]({\bf p}) J({\bf p}) \biggr\} \; , 
\end{eqnarray}
where the trace means
${\rm Tr} = \int \frac{d^{2} p}{(2 \pi)^{2}} \frac{d^{2} p'}{(2 \pi)^{2}} \; \delta({\bf p}-{\bf p}')$ 
and $G[\phi] = [G_{0}^{-1} + e \phi]^{-1}$.

Expanding the logarithm in the action with respect to the 
Coulomb field to quadratic order, 
the first two terms in Eq.~(\ref{eq:eff_action}) can be combined to 
the Gaussian action of the electric potential field
\begin{eqnarray}\label{eq:eff_gauss_action}
 S_{\rm F}[\phi]
  = - \frac{e^{2}}{2} \int_{\cal C} dt \int \frac{d^{2} q}{(2 \pi)^{2}} \; 
   \phi(-{\bf q}) V^{-1}({\bf q}) \phi({\bf q}) \; , 
\end{eqnarray} 
where 
\begin{eqnarray}\label{eq:v0_rpa}
 V^{-1}({\bf q}) = V_{0}^{-1}({\bf q}) + P_{0}({\bf q}) 
\end{eqnarray} 
is the dynamically screened interaction. 
Separating fields on the upper and lower parts of the Keldysh contour by 
introducing doublets, we can pass over to a convenient 
representation in terms of retarded and advanced functions 
(and mixtures thereof).
Then $P_{0}$ becomes a matrix in Keldysh space,
with a retarded part 
given by
\begin{eqnarray}\label{eq:polarization_rpa}
 P^{\rm R}_{0}({\bf q},\omega) 
 = \int \frac{d^{2} p}{(2 \pi)^{2}} \; 
  \frac{n_{\rm F}(\epsilon({\bf p}+{\bf q})) - n_{\rm F}(\epsilon({\bf p}))}
  {\omega + i \eta - \epsilon({\bf p}+{\bf q}) + \epsilon({\bf p})} \; ,
\end{eqnarray} 
where $n_{\rm F}(\epsilon) = [1+e^{\beta \epsilon}]^{-1}$ 
is the Fermi distribution function at inverse temperature $\beta$.
The Gaussian approximation made in deriving 
Eq.~(\ref{eq:eff_gauss_action}) is equivalent to the random phase 
approximation (RPA), which is well established for 
high electron densities 
$n_{s}$ corresponding to small values of the Brueckner parameter
$r_{s}<1$, but usually gives still very reasonable results for
larger $r_{s}$. 
As standard, $n_{s} = 1/\pi (r_{s} a_{0})^{2}$ with the 
effective Bohr radius $a_{0} = \varepsilon_{d}/m^{*}e^{2}$, where 
$\varepsilon_{d}$ is the dielectric constant.

The quantity of interest here is the DOS
\begin{eqnarray}\label{eq:DOS_def}
 \nu(\epsilon) = -\frac{2}{\pi} 
  \int \frac{d^{2} p}{(2 \pi)^{2}} \; {\rm Im} \, 
  G^{\rm R}({\bf p},\epsilon) \; ,
\end{eqnarray} 
where $G^{\rm R}$ is the retarded Green function.
The Keldysh matrix Green function is obtained as the mixed second
order functional derivative of the partition function with respect 
to the sources yielding 
\begin{eqnarray}
 G(r-r',t-t') = \int {\cal D}\phi \; G[\phi](r-r',t-t') \;
  e^{i S_{\rm F}[\phi]} \nonumber
\end{eqnarray} 
Following Schwinger \cite{Schwinger,Kopietz97} 
we try to find a functional $k[\phi](x,t)$ describing a 
local gauge transformation 
$\psi(x,t) \to e^{ik[\phi](x,t)} \psi(x,t)$ such that 
\begin{eqnarray}\label{eq:Schwinger_ansatz}
 && G[\phi](x-x',t-t') \nonumber \\
 && = e^{ik[\phi](x,t)} G_{0}(x-x',t-t') e^{-ik[\phi](x',t')} \; .
\end{eqnarray} 
Linearizing the dispersion relation near the Fermi surface, we can
write for $\omega \ll \epsilon_{\rm F}$ 
\begin{eqnarray}\label{eq:Eqn_of_motion}
 && [i \partial_{t} - \epsilon_{\rm F} + i {\bf v} \cdot \nabla 
  + e \phi(x,t)] G[\phi](x-x',t-t') \nonumber \\
 && = \delta(x-x',t-t')
\end{eqnarray}
where ${\bf v} = {\bf p}/m^{*}$ is the Fermi velocity. 
Eqs. (\ref{eq:Schwinger_ansatz}) and 
(\ref{eq:Eqn_of_motion}) 
determine $k$ as a linear functional of $\phi$. 
For the Green function 
$G^{>} = -i \langle \psi(x,t) \bar{\psi}(x',t') \rangle$
we then obtain 
\begin{eqnarray}
 G^{>}(x-x',t-t') = G^{>}_{0}(x-x',t-t') e^{J(x-x',t-t')} \nonumber
\end{eqnarray} 
where $G^{>}_{0}$ is the free Green function and
$J(x-x',t-t')$ is determined by the remaining
Gaussian path integral over the $\phi$ fields. This function, 
which we need for 
equal space arguments only, may be written as
\begin{eqnarray}\label{eq:j_of_t}
 J(t) \equiv J(x=x',t) = \int \frac{d\omega}{\pi} \; 
  \frac{e^{-i \omega t} - 1}{1 - e^{-\beta \omega}} \; 
   {\rm Im} \, Y(\omega) \; ,
\end{eqnarray} 
where
\begin{eqnarray}\label{eq:Yomega}
 Y(\omega) = - \int \frac{d^{2} q}{(2 \pi)^{2}} \; 
  \frac{1}{(\omega + i \eta - {\bf v} \cdot {\bf q})^{2}} \; 
  V^{\rm R}({\bf q},\omega) \; .
\end{eqnarray} 
Now, from Eq. (\ref{eq:DOS_def}) we finally obtain the formal result
\begin{eqnarray}\label{eq:DOS_P}
 \nu(\epsilon) = \nu_{0} \int d\epsilon' 
  \frac{1+e^{-\beta \epsilon}}{1+e^{-\beta \epsilon'}}
  P(\epsilon - \epsilon') \; ,
\end{eqnarray} 
where $\nu_{0}=m/\pi$ is the bare DOS and 
we introduced the spectral density 
\begin{eqnarray}\label{eq:P_of_E}
 P(\epsilon) = \int \frac{dt}{2 \pi} \; e^{i \epsilon t} 
  e^{J(t)} \; .
\end{eqnarray} 

To evaluate the DOS explicitly, we first note that the 
polarization function (\ref{eq:polarization_rpa}) reads for 
small ${\bf q}$ 
\begin{eqnarray}\label{eq:polarization_linearized}
 P^{\rm R}_{0}({\bf q},\omega) 
  = - \nu_{0} \int_{0}^{2 \pi} \frac{d\theta}{2 \pi} \; 
  \frac{vq\cos\theta}{\Omega-vq\cos\theta} 
\end{eqnarray} 
where $\Omega = \omega+i\eta$. 
If we introduce the function
\begin{eqnarray}
 g(q)=[1+\nu_{0} V_{0}(q)]^{-1/2} \; , \nonumber
\end{eqnarray} 
which is unity in the noninteracting case and less than one 
in presence of interactions, 
Eq.~(\ref{eq:Yomega}) 
may be combined with Eqs.~(\ref{eq:v0_rpa}) and 
(\ref{eq:polarization_linearized}) to read 
\begin{eqnarray}
 Y(\omega) &=& -\frac{1}{2 \pi} \int_{0}^{\infty} dq \nonumber \\ 
 &\times& \frac{\Omega q g^{2}(q) V_{0}(q)}{\left[\Omega^{2}-(vq)^{2}\right]
  \left[(\Omega+vq)\sqrt{\frac{\Omega-vq}{\Omega+vq}} 
   - [1-g^{2}(q)] \Omega \right]} \, . \nonumber 
\end{eqnarray}
The first factor in the denominator
represents the particle pole and the second the plasmon pole,
which is the solution $q_{\rm pl}(\omega)$ of the equation 
\begin{eqnarray}
 (vq)^{2} - [1-(1-g^{2}(q))^{2}] \omega^{2} = 0 \; . \nonumber 
\end{eqnarray}
As in 1D, the plasmon spectrum is gapless and makes an important
contribution to low energy properties.

In Eq.~(\ref{eq:j_of_t}) we need
the imaginary part
of $Y(\omega)$, which 
has a $\delta$--function contribution
at each of the poles of $Y(\omega)$, 
but also a regular part, and we find correspondingly
$ {\rm Im} \, Y = Y^{({\rm par})}
   + Y^{({\rm plas})} + Y^{({\rm reg})}$
with the particle and plasmon contributions (for $\omega>0$)  
\begin{eqnarray}
 Y^{({\rm par})}(\omega) = - \frac{1}{4 \nu_{0} v^{2}} \; ,\nonumber 
\end{eqnarray}
\begin{eqnarray}
 Y^{({\rm plas})}(\omega) &=& \frac{1}{2 \nu_{0} v^{2}} 
 \left[1 + \frac{q [1-g^{2}] g^{2} \nu_{0} \frac{d V_{0}}{dq}}
  {2-g^{2}} \right]^{-1}_{q=q_{\rm pl}(\omega)} \; , \nonumber 
\end{eqnarray}
and the regular part 
\begin{eqnarray}
 Y^{({\rm reg})}(\omega) 
 &=& - \frac{1}{4 \nu_{0} v^{2}} \int_{\omega/v}^{\infty} dq \;
  \nonumber \\
 && \hspace*{-2.5cm} \times 
  \frac{2 v^{2} q \omega (1-g^{2})}{ \pi \sqrt{(vq)^{2}-\omega^{2}} 
   \left[(vq)^{2} - [1-(1-g^{2})^{2}] \omega^{2} \right]} \; . \nonumber 
\end{eqnarray}

To proceed, we consider a bare interaction $V_{0}(q)$
of the form
\begin{eqnarray}
 V_{0}(q) = \frac{2 \pi e^{2}}{\varepsilon_{d} q} (1-e^{-2 q \Delta}),
  \nonumber
\end{eqnarray}
which is the 2D Coulomb interaction in presence of a 
screening gate at distance $\Delta$. Then 
${\rm Im} \, Y(\omega)$ contains three energy scales: 
$\omega_{0} = 2 \pi \nu_{0} v^{2} = 4 \epsilon_{\rm F}$, 
$\Omega_{\kappa} = \kappa v$,
where $\kappa = 2 \pi \nu_{0} e^{2}/\varepsilon_{d}$ is the 
two--dimensional inverse screening length, and 
$\Omega_{\Delta} = v/2\Delta$. For the bare 2D Coulomb interaction, i.e. \
$\Delta \to \infty$, only the two scales 
$\omega_{0}$ and $\Omega_{\kappa}$ remain, that are related by
$r_{s} = 2 \Omega_{\kappa}/\omega_{0}$. 
For finite $\Delta$ it is convenient to introduce the ratio 
$\lambda = \Omega_{\kappa}/\Omega_{\Delta}= 2 \kappa \Delta$.

\begin{figure}[b]
\begin{center}
 \leavevmode
  \epsfxsize=0.365 \textwidth
  \epsfbox{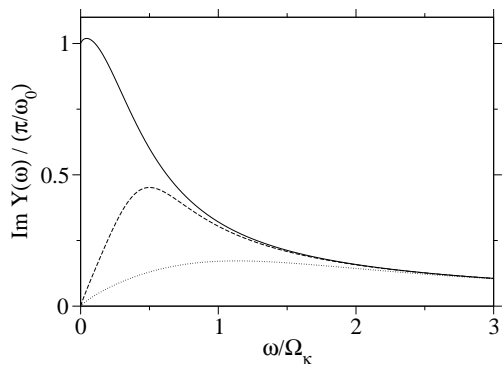}
  \vspace*{-.0cm}
\caption{\label{fig:Y_omega} The function ${\rm Im} \, Y(\omega)$
  for an unscreened bare Coulomb potential (solid curve) and in
  the presence of a gate at distance $\Delta$ with 
  $\kappa \Delta = 5$ (dashed curve) and $\kappa \Delta = 1$
  (dotted curve). $Y$ in units of $\pi/\omega_{0}=1/2 \nu_{0} v^{2}$, 
  $\omega$ in units of $\Omega_{\kappa}=\kappa v$.}
\end{center}
\vspace*{-0.5cm}
\end{figure}

In 2D disordered systems ${\rm Im} \, Y$ diverges at low frequencies
which leads to a divergence of $J(t)$ for $t \to \infty$ implying
a total suppression of $\nu(\epsilon)$ for $\epsilon \to 0$. The 
same is true for 1D ballistic wires. In the 2D ballistic case
considered here, ${\rm Im} \, Y$ remains
finite or even vanishes in presence of a gate as shown in 
Fig.~\ref{fig:Y_omega}.
Accordingly, at $T=0$, $J(t)$ approaches a constant $J_{\rm as}$ for
$t \to \infty$, which is given by 
\begin{eqnarray}\label{eq:J_asymptotic}
 J_{\rm as} = - \int_{0}^{\infty} \frac{d\omega}{\pi} \; 
  {\rm Im} \, Y(\omega).
\end{eqnarray}
It is important to note that the explicit result for 
${\rm Im} \, Y(\omega)$ given above does not suffice to 
determine $J_{\rm as}$
explicitly, since the integral (\ref{eq:J_asymptotic}) 
also has contributions from frequencies that are not small compared to 
$\epsilon_{\rm F}$ where the true parabolic dispersion must be used,
which leads to a faster decay of ${\rm Im} \, Y(\omega)$ at high
frequencies.
However, this high energy behavior chiefly affects the quantity
$J_{\rm as}$, while for large times the remaining part $J'(t)$ 
in the decomposition
\begin{eqnarray}
 J(t) = J_{\rm as} + J'(t) \nonumber 
\end{eqnarray}
is determined by the low energy behavior of $Y(\omega)$.

Accordingly, the spectral density (\ref{eq:P_of_E}) splits into 
\begin{eqnarray}
P(E) = S \delta(E) + P'(E) \; , \nonumber
\end{eqnarray}
with the regular part 
\begin{eqnarray}
  P'(E) := S \theta(E)
  \int \frac{dt}{2 \pi} \; e^{i E t} \; (e^{J'(t)}-1) \; , \nonumber
\end{eqnarray}
where $S = e^{J_{\rm as}}<1$ is a suppression factor. 
Hence, the $\delta(E)$--function form of the spectral density
of a noninteracting system partially survives in 2D ballistic
electron systems. 

From Eq.~(\ref{eq:DOS_P}) the DOS now reads
\begin{eqnarray}
 \nu(\epsilon) = S \nu_{0} + \nu'(\epsilon) \; , \nonumber
\end{eqnarray}
where the energy dependent part is given by 
\begin{eqnarray}
 \nu'(\epsilon) = \nu_{0} \int_{0}^{|\epsilon|} dE \; P'(E) \; . \nonumber
\end{eqnarray}
Clearly, $\nu(\epsilon)$ has a nonzero value $\nu(0) = S \nu_{0}$
at $\epsilon=0$, 
which depends on the integral property $J_{\rm as}$ and thus 
on the interaction strength as opposed to earlier 
predictions \cite{Khveshchenko98,MA01}. Also $\nu'(\epsilon)$
contains the suppression factor $S$ 
as a prefactor.
Further, the slope and curvature 
of $\nu(\epsilon)$ at $\epsilon=0$ are not universal but depend
on the interaction strength $r_{s}$.

\begin{figure}[btp]
\begin{center}
 \leavevmode
  \epsfxsize=0.38 \textwidth
  \epsfbox{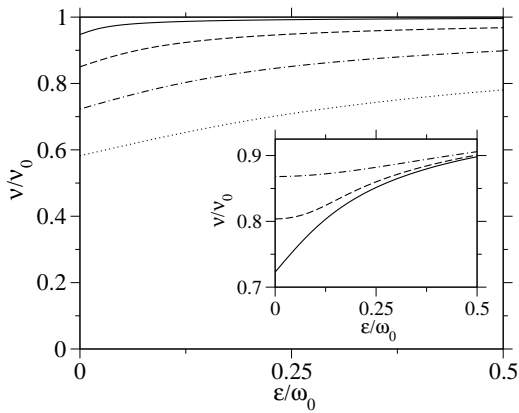}
  \vspace*{-.0cm}
\caption{\label{fig:nu} Density of states for $r_{s}=0.1,0.3,0.6,1.0$ 
  (topmost to lowest graph). Inset: 
  The correction to the DOS for $r_{s}=0.6$
  without a gate (solid curve) is reduced by a gate at 
  distances $\Delta$ with $\kappa \Delta = 5$ (dashed curve) 
  and $\kappa \Delta = 1$ (chain--dotted curve).}
\end{center}
\vspace{-0.6cm}
\end{figure}

To determine the value of $J_{\rm as}$ and $S$ explicitly without artificial 
cutoffs, we either have to use the full quadratic dispersion or
to approximate the less important regular part of $Y(\omega)$. 
Since $Y^{({\rm reg})}(\omega)$ approaches $-1/4 \nu_{0} v^{2}$
for small and large frequencies and does not deviate much from
this limiting value in between, we replace it by that constant. 
In this approximation, which does not necessarily capture the 
true high frequency behavior and thus gives only an estimate of
$J_{\rm as}$ when inserted into Eq.~(\ref{eq:J_asymptotic}), the 
function ${\rm Im} Y(\omega)$ is of the form
${\rm Im} Y(\omega)=\frac{\pi}{\omega_{0}} y(\omega/\Omega_{\kappa},\lambda)$. 
From this scaling form we find 
$J_{\rm as} = - r_{s} \zeta(\lambda)$ with 
$\zeta(\lambda)=1/2 \int_{0}^{\infty} du \; y(u,\lambda)$, 
which indicates that the 
factor $S=e^{-r_{s} \zeta(\lambda)}$ rapidly suppresses $\nu(0)$
with increasing $r_{s}$. With decreasing gate distances the 
suppression factor approaches $1$, which means that interaction effects 
become weaker.
These features of the DOS are also apparent from Fig.~\ref{fig:nu}
which displays results for various values of $r_{s}$ with and
without gate. Without a gate the DOS is cusplike
near $\epsilon=0$, whereas in presence of a gate the slope at 
$\epsilon=0$ vanishes. This latter result is in accordance with
the perturbative analysis \cite{MA01}.

For finite temperatures and in the absence of a gate the denominator 
in Eq.~(\ref{eq:j_of_t}) behaves singular at $\omega=0$.
While $|J(t)|$ then exceeds $|J_{\rm as}|$, 
it turns out that $\nu(0)$ always increases if the temperature is raised. 
This means that finite temperatures smear out the cusplike 
DOS and reduces the size of interaction effects near $\epsilon=0$. 

Now, two remarks are in order. Strongly interacting semiconductor
systems, where values for $r_{s}$ up to 40 have been observed
\cite{Sarachik01}, are usually disordered, but a ballistic regime is 
feasible \cite{Proskuryakov02}. For large $r_{s}$ our analysis 
does not necessarily apply and, in particular, as was discussed 
above, we cannot obtain reliable values for $J_{\rm as}$.
However, if no scales other than $\epsilon_{\rm F}$ and 
$\Omega_{\kappa}$ enter, $J_{\rm as}$ should be of the form
$J_{\rm as} = -r_{s} \zeta(r_{s},\lambda)$. 
Then $S$ will vanish for $r_{s} \gg 1$, provided 
$\lim_{r_{s} \to \infty} r_{s} \zeta(r_{s},\lambda) = \infty$. 

Second, in the absence of a gate we have 
$\nu(\epsilon) \approx \nu_{0} S [1 + 
|\epsilon|/4 \epsilon_{\rm F} + {\cal O}(\epsilon^{2})]$. 
This differs from previous perturbative approaches giving
$\nu(\epsilon) \approx \nu_{0} (1 + |\epsilon|/4 \epsilon_{\rm F})$ 
\cite{Khveshchenko98,MA01} independent of $r_{s}$. The lack of
a dependence on the interaction strength can be traced
back to the fact that in these works ${\rm Im} \, Y(\omega)$ is 
effectively replaced by a constant and cut off at the Fermi level. 
Then, no interaction scale remains and therefore the authors obtain 
a universal (interaction independent) result.
From Fig.~\ref{fig:Y_omega} we see that this is not the case, 
but the dominant plasmon contribution to ${\rm Im} \, Y(\omega)$ 
falls off on the scale $\Omega_{\kappa}$, which
depends on the interaction strength. Indeed, if we turn off 
the interaction, the corrections to the noninteracting DOS $\nu_{0}$
vanish. 

Throughout this Letter, we have considered a pure 2DEG without 
disorder. In weakly disordered 2DEGs with elastic 
mean time $\tau$ electrons move diffusively on long time scales
corresponding to excitation energies small compared to $1/\tau$. 
Then, at $T=0$ the DOS drops down to zero for energies
below $1/\tau$ where the behavior passes over to 
an Altshuler--Aronov ZBA 
[3--6].
Some authors have investigated the crossover from the 
diffusive to the quasiballistic limit perturbatively 
\cite{RAG97,Khveshchenko98,MA01}. The 
clean limit of these treatments does not yield our results for
the reasons discussed above.

Experimentally it is possible to realize very clean and strongly
interacting 2DEGs (or 2D hole gases) 
\cite{Sarachik01,Proskuryakov02}, but so far mainly 
transport measurements have been performed in the context of 
the metal--insulator transition. Tunneling experiments in 
samples with high $r_{s}$ have been suggested \cite{Sarachik01}, 
but have not yet been performed, since they are more involved, 
because a counter electrode has to be added at a small distance  
without influencing the quality of the device. Nevertheless, 
tunneling experiments with a ballistic 2DEG seem to be feasible
in the future.
Interaction effects on the DOS in 2DEGs are also relevant for 
experiments with 2D--1D tunnel junctions \cite{dePicciotto00},
because there not only the DOS of the 1D wire is measured, but
a convolution with the suppressed DOS of the 2DEG.


We would like to thank R. Egger and M. Sarachik for
helpful correspondence.
Financial support was provided by the DFG.


\end{document}